\documentclass{article}

\usepackage{arxiv}

\usepackage[utf8]{inputenc} 
\usepackage[T1]{fontenc}    
\usepackage{hyperref}       
\usepackage{url}            
\usepackage{booktabs}       
\usepackage{amsfonts}       
\usepackage{nicefrac}       
\usepackage{microtype}      
\usepackage{lipsum}		
\usepackage{graphicx}
\usepackage{natbib}
\usepackage{doi}
\usepackage{natbib}
\usepackage{arxiv}
\usepackage{subcaption} 
\usepackage[utf8]{inputenc} 
\usepackage[T1]{fontenc}    
\usepackage{hyperref}       
\usepackage{url}            
\usepackage{booktabs}       
\usepackage{amsfonts}       
\usepackage{nicefrac}       
\usepackage{microtype}      
\usepackage{lipsum}
\usepackage{graphicx}
\graphicspath{ {./images/} }
\usepackage{amsmath}
\usepackage{braket}
\usepackage{algorithm}       
\usepackage{algpseudocode}

\usepackage[utf8]{inputenc} 
\usepackage[T1]{fontenc}    
\usepackage{hyperref}       
\usepackage{url}            
\usepackage{booktabs}       
\usepackage{amsfonts}       
\usepackage{nicefrac}       
\usepackage{microtype}      
\usepackage{hyperref}

\usepackage{amsmath}
\usepackage{amssymb}
\usepackage{epstopdf}
\title{A hybrid quantum solver for the Lorenz system}

\author{
  Sajad Fathi Hafshejani$^1$ 
  \And
  Daya Gaur$^1$
  \And
  Arundhati Dasgupta$^2$ 
  \And
  Robert Benkoczi$^1$
  \And
  Narasimha Gosala$^2$
  \And
  Alfredo Iorio$^{3,4,5}$ 
  \\ 
  $^1$Department of Math and Computer Science, University of Lethbridge, Lethbridge, AB T1K 3M4, Canada;\\ sajad.fathihafshejan@uleth.ca(S.F.H.), daya.gaur@uleth.ca(D.G.),  robert.benkoczi@uleth.ca(R.B.)\\
  $^2$The Department of Physics \& Astronomy, University of Lethbridge, Lethbridge, AB T1K 3M4, Canada; \\
  arundhati.dasgupta@uleth.ca(A.D.), narasimha.gosala@uleth.ca(N.G.)\\
 $^3$ Institute of Particle and Nuclear Physics, Faculty of Mathematics and Physics, Charles University,\\ V Holešovičkách 2, 180 00 Prague 8 -- Czech Republic; alfredo.iorio@mff.cuni.cz(A.I.)\\
  $^4$Democritos Technologies, Rybna 24, 110 00 Prague 1 -- Czech Republic\\
  $^5$ Department of Physics, University of Calabria, 87036 Rende (CS) -- Italy
}

\begin{document}
\maketitle
\begin{abstract}
We develop a hybrid classical-quantum method for solving the Lorenz system. We use the forward Euler method to discretize the system in time, transforming it into a system of equations. This set of equations is solved using the  Variational Quantum Linear Solver (VQLS) algorithm. We present numerical results comparing the hybrid method with the classical approach for solving the Lorenz system. 
The simulation results demonstrate that the VQLS method can effectively compute solutions comparable to classical methods. The method is easily extended to solving similar nonlinear differential equations.
\end{abstract}


\section{Introduction}
Dynamical systems theory, along with its numerical and simulated counterparts, has been employed to study various phenomena such as weather patterns, population dynamics, economic trends, the flow of chemical and biological elements, \textit{etc} \cite{thompson2002nonlinear}. Whether it is the macroscopic dynamics of a temperature-pressure system or the microscopic complexities of a nonlinear system with bifurcations and irreversible physics, mathematical modeling, based on dynamical system theory, offers a valuable tool for analysis. 

Dynamical systems may develop deterministic chaos, the Lorenz system being a noticeable example. It was introduced by Lorenz in 1963 \cite{lorenz1963deterministic}
 as a simple \textit{nonlinear} model of heat convection, and it stands out as one of the earliest attempts to capture atmospheric physics through a model consisting of three differential equations. It is a chaotic three-dimensional system that has been studied extensively, mostly numerically. For a known set of parameter values, the three-dimensional motion converges to the well-recognized butterfly-shaped attractor, which can be observed by solving the system numerically. However, no analytic proof of this fact is known. 

In this paper, we focus on how quantum computing might help to improve the numerical solutions of this important nonlinear dynamical system. Unlike classical physics, quantum mechanics generally only provides probabilities for the outcomes of measurements. Quantum systems evolve in a linear fashion and when a quantum system in superposition is measured, the act of measurement forces the system into a particular state, and the theory predicts the probabilities of obtaining various possible outcomes. This inherent probabilistic measurement outcomes is a crucial feature of quantum uncertainty as opposed to uncertainty in a weather system which arise due to nonlinear behaviour \cite{penrose2011uncertainty}. 

When we linearize the system using the forward Euler method and solve it on a classical computer, the subsequent point is uniquely determined from the current point and is obtained by solving a system of linear equations. Classical chaos arises in systems governed by nonlinear equations with a positive Lyapunov coefficient, where small variations in the initial conditions can lead to drastically different outcomes over time. This sensitivity to initial conditions is visible in solutions computed on a classical computer using the forward Euler method. 

{When solving a single iteration of the Euler method using a quantum algorithm, there are two sources of uncertainty. The initial state can only be prepared with a certain level of precision, and the solution computed by a quantum algorithm is inherently probabilistic, resulting in a random sample rather than a deterministic value. However, it is important to note that this randomness does not introduce new features or chaotic behavior into the system. The differences in trajectories between classical and quantum solutions arise from the inherent probabilistic nature of quantum algorithms, not from any additional characteristics of the Lorenz system itself. Both methods—classical and quantum—can provide approximate solutions, but the quantum method may do so more efficiently by exploiting quantum randomness, similar to how stochastic classical methods use randomness.

Suppose we replace the equation-solving step in the Euler method with a quantum algorithm; given the new source of uncertainty, the following questions arise: i) does the chaos in the system increase when a quantum subroutine is used in each iteration? ii) does the system still have the butterfly attractor (for the choice of parameters given above)? We answer these questions in the affirmative under the assumption that it is possible to prepare the state exactly and recover the solution exactly. These assumptions can only be justified with sufficient advances in Quantum random-access memory (QRAM) and quantum tomography techniques \cite{lanyon2017efficient,cramer2010efficient}

One of the earliest attempts, to solve nonlinear differential equations using a quantum approach, is due to \citet{leyton2008quantum}. They use the forward Euler method and multiple copies of an initial state, which are evolved according to the \textit{linear} Euler system. This algorithm scaled poorly as a function of the time step, as shown in \citet{berry2014high}, which goes on to show that a far more efficient quantum procedure can be obtained for linear differential equations. For this, they compute the bounds on the condition number of the matrix. \citet{liu2021efficient} gives an efficient quantum algorithm for dissipative nonlinear differential equations. 

\citet{berry2017quantum} gives an exponentially improved quantum algorithm for solving linear ordinary differential equations with constant coefficients. 
Although the paper primarily focuses on diagonalizable matrices, it can be extended to approximate solutions for non-diagonalizable matrices by using nearby diagonalizable matrices, as discussed in \cite{krovi2023improved}.

\citet{krovi2023improved} presents a quantum algorithm for solving linear, inhomogeneous ordinary differential equations (ODEs). The algorithm shows improved gate and query complexity for specific diagonalizable classes of matrices and is extended to handle non-diagonalizable and singular matrices. 

\citet{lloyd2020quantum} introduced a quantum algorithm called the ``quantum nonlinear solver'' 
which can solve nonlinear differential equations – a critical component of weather prediction models. 
Applications of these algorithms to weather prediction have been studied by \citet{tennie2023quantum}. This solver leverages multiple copies of a quantum state to simulate nonlinearity, potentially providing an advantage over classical methods. Initial tests of this algorithm on a simplified model have shown promising results, demonstrating its agreement with classical methods and ensemble averages for a single-particle system, $dx/dt = x - \alpha x^3$. The algorithm discretizes the time domain, similar to the Forward Euler method. The set of equations can be represented in matrix form as $Ax = b$, with the state vector $x$ obtained 
using the Harrow-Hassidim-Lloyd (HHL) algorithm. The nonlinear forward Euler method is utilized in the case of nonlinear differential equations. 

For the potential applications of quantum computing to climate change studies, see the recent review by \citet{rahman2024climate}. The authors have identified quantum principal component analysis (qPCA) and HHL as algorithms with quantum advantages that are crucial for analyzing climate models. It is difficult to leverage the advantages of quantum computing for large weather and climate datasets due to limited readout capacity and data accessibility challenges \cite{tennie2023quantum}. 

\citet{armaos2024quantum} explored the use of the Variational Quantum Eigensolver (VQE), a quantum computing algorithm, to analyze the Lorenz system. 
The main focus of their analysis was to determine the eigenvalues of the modified Jacobian matrix using VQE. This identified stable and unstable points of the Lorenz system .

{\sc Contributions:} The main contribution of this work is a method, described in Subsection \ref{sec:method}, for solving the Lorenz system of atmospheric convection. This method uses a hybrid quantum-classical algorithm. The technique involves discretizing the Lorenz system in time using the forward Euler method. The resulting system of equations is solved using the Variational Quantum Linear Solver (VQLS) described in Section \ref{sec:vqls}. Section \ref{sec:algorithm} is dedicated to the algorithm. The results of the simulations are reported in Section \ref{sec:simul}. The simulation of the VQLS method is computationally intensive on classical computers compared to classical algorithms that require inverting a matrix. However, it requires significantly fewer qubits (three) compared to other quantum algorithms like the HHL algorithm, which requires a minimum of nine qubits for one iteration, making it potentially suitable for implementation on near-term Noisy Intermediate-Scale Quantum (NISQ) computers. Simulations demonstrate that the VQLS method effectively computes solutions comparable to classical methods. See Figure \ref{fig:clas-vqls} for an attractor discovered using both the quantum and classical methods proposed in Subsection \ref {sec:method}. A detailed error analysis of the two methods; classical and quantum based on Richardson extrapolation is in Subsection \ref{sec:error}. Unlike previous quantum approaches \cite{tennie2023quantum}, for computing atmospheric dynamics,  that used multiple copies of the state, this method simulates only a single time step, avoiding measurement errors that accumulate over time. The method developed in this paper is easily extended to solve other similar nonlinear differential equations.  Results and limitations of the approach are discussed in Subsection \ref{sec:results}.

\section{ Lorenz system }

The Lorenz system consists of three coupled, nonlinear ordinary differential equations. The system was first introduced by Edward Lorenz in 1963 in a paper where he explored the underlying mechanisms of long-range weather prediction \cite{lorenz1963deterministic}.
The equations describe the rate of change of three quantities over time, often interpreted physically as the movement of a fluid cell in a larger circulation. The Lorenz equations are  as follows:

   \begin{eqnarray}
\frac{dx}{dt} &=& \sigma(y - x) \,,\label{lorenzx} \\
\frac{dy}{dt} &=& x(\rho - z) - y \,, \label{lorenzy} \\
\frac{dz}{dt} &=& xy - \beta z \,,\label{lorenzz}
   \end{eqnarray}
where
\begin{itemize}
\item \(x\), \(y\), and \(z\) represent the state variables of the system. These variables can be thought of as proportional to the intensity of the convective motion, the temperature difference between ascending and descending currents, and the deviation of the vertical temperature profile from linearity, respectively.
\item \( \sigma \) is the Prandtl number, representing the ratio of viscous diffusivity to thermal diffusivity.
\item \( \rho \) represents the Rayleigh number, which measures the thermal buoyancy force relative to viscous damping in the fluid.
 \item \( \beta \) is a geometric factor associated with the problem.
\end{itemize}
For certain values of the parameters, $\sigma=10, \rho=28,  \beta=8/3$, the three-dimensional motion converges to the well-known butterfly-shaped attractor, which can be observed by solving the system numerically.

It is customary to use Lyapunov exponents to measure how quickly two initially close points in a system diverge from each other, as time progresses, indicating the rate of separation of nearby trajectories. Positive Lyapunov exponents suggest chaotic behaviour and sensitivity to initial conditions, while negative or zero exponents indicate stability or convergence of trajectories over time. The existence of positive Lyapunov exponents in 
the Lorenz system shows the presence of chaos \cite{thompson2002nonlinear, sparrowlorenz}, and indeed the Lorenz system can be chaotic, as shown in Figure \ref{fig:chaotic}. 


These dynamical systems are also studied for the stability behaviour, the existence of attractor points, saddle points, etc, under the time flow. As systems become increasingly complex, they require more computational power. Quantum computers, a technology of the future, promise more efficient and faster ways to solve numerical systems and speed up numerical methods for solving such differential equations.

\begin{figure*}[t!]
    \centering
    \begin{subfigure}[t]{0.45\textwidth}
        \centering
        \includegraphics[height=2in]{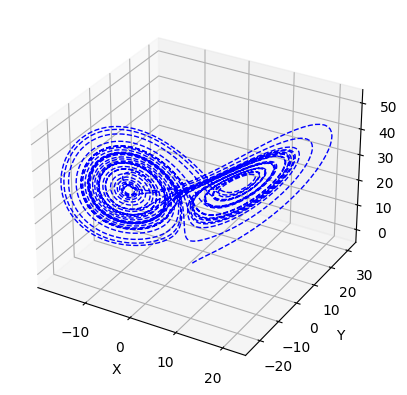}
        \caption{The starting point is $(1,2,-4)$.}
    \end{subfigure}%
    ~ 
    \begin{subfigure}[t]{0.45\textwidth}
        \centering
        \includegraphics[height=2in]{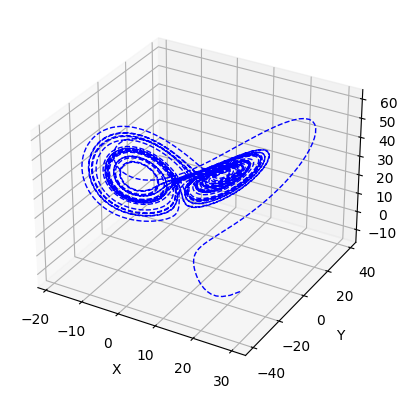}
        
        \caption{The starting point is $(30,-40,10)$. }
    \end{subfigure}
    
    \caption{The trajectory was generated using the method described in Subsection \ref{sec:method} on a classical computer.}
    \label{fig:chaotic}
\end{figure*}



A chaotic system exhibits complex and unpredictable behaviour.
Figure \ref{fig:start} shows two very different trajectories for two start points that are extremely close to each other $(1e-16,1e-16,1e-16)$ and $(1e-16,-1e-16,1e-16)$. 
This means that small changes in the starting point can lead to significantly different outcomes, making long-term prediction impossible. 
The Lorenz system is a classic example of such a system, displaying chaotic solutions for specific parameter values and initial conditions  \cite{lorenz1963deterministic,palmer2014real}. This system has been extensively analyzed in the literature and is a fundamental example in the study of chaotic systems \cite{thompson2002nonlinear, sparrowlorenz}.

\begin{figure*}[!ht]
\centering
  \includegraphics[width=0.4\textwidth]{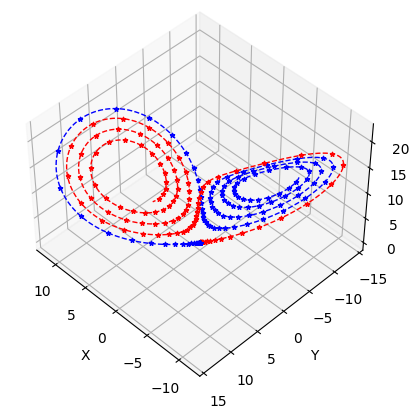}
  \caption{Blue: the starting point is $(1e-16,1e-16,1e-16)$. Red: the starting point is $(1e-16,-1e-16,1e-16)$ and the parameters are $(13.92655741, 10, 8/3)$. The two trajectories generated using the method in Section \ref{sec:method} differ widely. }
  \label{fig:start}
\end{figure*}

\subsection{Linearization}

Consider the following approximation of the Lorenz system, (\ref{lorenzx})-(\ref{lorenzz}), obtained by dropping the nonlinear terms in Eq.(\ref{lorenzy}), $xz$, and in Eq.(\ref{lorenzz}), $xy$
\begin{align}
\dot{x} & = \sigma (y -x ) \,, \\
\dot{y} & = \rho x - y \,, \\
\dot{z} & = -\beta z \,,
\end{align}
where, we denote $dx/dt$ as $\dot{x}$, etc.

Now, we can use the forward Euler method to discretize this system in time.  If $\dot{x} = (x_{n+1} - x_n)/h$, etc, for small $h$, the linearized system becomes
\begin{align}
x_{n+1} & = x_n + h \sigma (y_n -x_n ) \,, \\
y_{n+1} & = y_n + h\rho x_n - hy_n \,, \\
z_{n+1} & = z_n - h\beta z_n \,.
\end{align}
By denoting with $w_n = (x_n, y_n, z_n)$ the 3-dimensional vector, then, for all $n$, the linearized system can be written as 
\begin{equation} \label{linearmatrixeq}
w_{n+1} = A_L \, w_n  \,,
\end{equation}
where ``$L$'' stands for \textit{linear}
\begin{equation}
A_L = \begin{bmatrix}
1 - h \sigma  & h \sigma & 0 \\
h \rho & 1-h & 0 \\
0 & 0 & 1 - h\beta
\end{bmatrix} \,,
    \end{equation}
is the $3 \times 3$ matrix encoding the single time step, from $n$ to $n+1$, of the linear evolution.

If the system evolves for $T$ time steps, the evolution is described by the following matrix equation 
\begin{equation} \label{linearmatrixeqB}
B_L \, w \equiv
\begin{bmatrix}
I & \mathcal{O}    & \ldots & \mathcal{O} & \mathcal{O} \\
A_L & -I & \ldots & \mathcal{O} & \mathcal{O} \\
\vdots   & \vdots & \ddots &\vdots &\vdots  \\
\mathcal{O} & \mathcal{O}    &  \ldots & -I & \mathcal{O} \\
\mathcal{O} & \mathcal{O}    & \cdots & A_L & -I 
\end{bmatrix}
\begin{bmatrix}
w_1 \\
w_2 \\
\vdots \\
w_{T-1} \\ 
w_T
\end{bmatrix}
= 
\begin{bmatrix}
w_1 \\
O \\
\vdots \\
O \\
O
\end{bmatrix} \equiv b \,,
\end{equation}
where $B_L$ is a $3 T \times 3 T$ matrix, and $w = (w_1, ..., w_T)$ is a $3T$-dimensional vector, $I$ is the $3\times 3$ identity matrix, whereas $\mathcal{O}$ is a $3\times 3$ matrix with all $0$ entries, $O$ is a 3-dimensional vector with all $0$ entries, hence $b = (w_1, O, ...,O)$ is also a $3T$-dimensional vector. 

Indeed, from the first line we get $w_1 = w_1$, but then, from the second we obtain $A_L w_1 - w_2 = O$, i.e., $w_2 = A_L w_1$, that is the matrix equation (\ref{linearmatrixeq}) for $n=1$, and so on, till the last line that gives $A_L w_{T-1} - w_T = O$, i.e., $w_T = A_L w_{T-1}$, that is the matrix equation (\ref{linearmatrixeq}) for $n=T-1$. 

The evolution of the system is given by $w = (w_1, w_2, \ldots, w_T)$ which can be obtained the inverting the block diagonal matrix $B_L$ and multiplying it with $b$, i.e., $w = B_L^{-1} b$. 

We can also use HHL to compute the solution of the system (\ref{linearmatrixeqB}). This linearization on a collection of qubits along with HHL is used in \cite{tennie2023quantum} for solving atmospheric dynamics using quantum approaches. The advantage offered is the exponential speedup due to HHL subject to limitations posed by the process of initial state preparation, solution read out and noisy quantum bits constraints. A similar linearization is used in \cite{armaos2024quantum} along with a variational quantum eigensolver to study the stability of a simplified model of the atmospheric system (Lorenz system) at equilibrium points.

\subsection{Inclusion of nonlinear terms}
\label{sec:method}

When the nonlinear terms are disregarded, the resulting system yields to a simpler and more tractable analysis. However, the linear system has important limitations:
\begin{itemize}
    \item It is unable to capture global behavior, especially in regions far from equilibrium.
    \item Chaotic behavior, bifurcations, and other complex dynamics are completely overlooked by linear analysis. In particular, it does not encompass chaos, which is a defining feature of the Lorenz system. 
    \item The effectiveness of the linear approximation can vary significantly with changes in parameters (\(\sigma\), \(\rho\), \(\beta\)). Some dynamics observable in one set of parameter values may be absent in others.
\end{itemize}



To overcome these challenges, we propose here a new method incorporating nonlinear terms. The discretized version of the nonlinear Lorenz system (\ref{lorenzx}) - (\ref{lorenzz}), obtained through the very same steps as for the linear case, is simply
  \begin{eqnarray}
x_{n+1} &=& x_n + h \sigma (y_n - x_n) \,, \label{system2x} \\
y_{n+1} &=& y_n + h [x_n (\rho - z_n) - y_n] \,, \label{system2y}\\
z_{n+1} &=& z_n + h (x_n y_n - \beta z_n) \,. \label{system2z}
  \end{eqnarray}
The value \( h \) is the timestep that controls the resolution of the simulation, e.g., $h=10^{-5}$. It is important to handle the timestep \( h \) carefully to ensure stability and accuracy.

We rewrite this system as
  \begin{eqnarray}
x_{n+1} - h \sigma y_n - x_n (1- h \sigma) &=& 0 \label{system3x} \\
y_{n+1} - h x_n (\rho - z_n)  -y_n(1-h) &=& 0 \label{system3y} \\
z_{n+1} - h x_n y_n - z_n (1-\beta h) &=& 0.\label{system3z}
  \end{eqnarray}
that in matrix form is 
\begin{equation} \label{matrixnonlineq}
A_{NL} W = b_{NL}    
\end{equation}
where  ``$NL$'' stands for \textit{nonlinear}, $A_{NL}$ is an $8 \times 8$ matrix, that we shall explicitly write in a moment,  $W$ and $b_{NL}$ are 8-dimensional vectors given by 
\begin{equation}
W = (x_n, y_n, z_n,x_{n+1}, y_{n+1}, z_{n+1},x_n z_n, x_n y_n)  \,,    
\end{equation}
and 
\begin{equation} \label{bNL}
    b_{NL} = (x_n, y_n, z_n, 0, 0, 0, x_n z_n, x_n y_n) \,.
\end{equation}
The vector $W$ contains the solutions.

It is important to notice that the matrix equation (\ref{matrixnonlineq}) refers to a single time step, henceforth it is the nonlinear generalization of Eq. (\ref{linearmatrixeq}), and not of Eq. (\ref{linearmatrixeqB}), that refers to multiple time steps. With the latter, it shares the higher dimensionality, as compared to the linear system of Eq. (\ref{linearmatrixeq}) (8 dimensions vs 3 dimensions), but that is due to the inclusion of the nonlinear terms into a single time step.

Thus, system (\ref{system3x})-(\ref{system3z}), in its matrix form (\ref{matrixnonlineq}), can be written as
\begin{eqnarray}\label{final_system}
A_{NL} W \equiv
\begin{bmatrix}
1 & 0 & 0 & 0 & 0 & 0 & 0 & 0 \\
0 & 1 & 0 & 0 & 0 & 0 & 0 & 0 \\
0 & 0 & 1 & 0 & 0 & 0 & 0 & 0 \\
-(1-h \sigma) & -h \sigma & 0 & 1 & 0 & 0 & 0 & 0\\
-h\rho &-(1-h) & 0 & 0 & 1 & 0 & h & 0\\
0 & 0 & -(1-\beta h) & 0 & 0 & 1 & 0 & -h \\
0 & 0 & 0 & 0 & 0 & 0 & 1 & 0 \\
0 & 0 & 0 & 0 & 0 & 0 & 0 & 1 \\
\end{bmatrix}
\begin{bmatrix}
x_n\\
y_n \\
z_n \\
x_{n+1} \\
y_{n+1} \\
z_{n+1} \\
x_n z_n \\
x_n y_n
\end{bmatrix} =
\begin{bmatrix}
x_{n} \\
y_{n} \\
z_{n} \\
0 \\
0 \\
0 \\
x_n z_n \\
x_n y_n
\end{bmatrix} \equiv b_{NL} \,.
\end{eqnarray}

The solution $x_{n+1}, y_{n+1}, z_{n+1}$ is obtained by inverting the matrix $A_{NL}$ and multiplying it with $b_{NL}$. 

This inversion can be done classically or using HHL \cite{lloyd2020quantum} as in \cite{tennie2023quantum}.  HHL assumes Hermitian matrix, therefore one has to solve for {${\mathcal A}_{NL} {\mathcal W}_{NL} = B_{NL}$ where ${\mathcal A}_{NL} = \begin{bmatrix} \mathcal{O}_8 &A_{NL} \\   A_{NL}^\dagger &\mathcal{O}_8\end{bmatrix}$, ${\mathcal W}_{NL} = (\mathbf{O}_8,W)$ and $B_{NL} = (b_{NL}, \mathbf{O}_8)$ and with $\mathcal{O}_8$ a $8\times 8$ matrix with all $0$ entries, and $\mathbf{O}_8$ is an 8-dimensional column vector of zeros.}
We need $4$ qubits to represent the input $b_{NL}$, one qubit for controlled rotation, and say 4 qubits for the clock register (needed for simulation of the Hamiltonian). So, a total of 9 qubits are needed if HHL is used to solve $A_{NL} W = b_{NL}$. The precision increases with the increase in the number of clock qubits. The condition  number for $\mathcal{A}_{NL}$ is 3.03 for $h$, step size,  of $0.01$.
    \\ {Although the HHL and VQLS methods are both prominent algorithms for solving linear systems, they present distinct advantages and limitations. The HHL algorithm is known for its theoretical exponential speedup over classical methods for certain classes of linear systems, particularly those matrices $A$ with the sparse and well-conditioned \cite{Harrow2009, Childs2017}. However, HHL is highly susceptible to quantum gate errors and qubit decoherence, which significantly impair its performance on current quantum hardware \cite{Campbell2017, Preskill2018}. Additionally the simulation of HHL is computationally intensive.
\\
In contrast, the VQLS method employs simpler and often shorter quantum circuits, making it more feasible on available noisy intermediate-scale quantum (NISQ) hardware \cite{bravo2023variational, Cerezo2021}. While VQLS may not achieve the same exponential speedup as HHL, its variational approach facilitates noise mitigation, resulting in greater robustness and reliability in noisy quantum environments \cite{Endo2018, McArdle2020}. Additionally, VQLS requires fewer quantum resources by combining quantum subroutines with classical optimization, making it particularly suited to NISQ devices, where quantum coherence time is limited \cite{McClean2017, Benedetti2019}.
 }

Here, we should emphasize again that, while this work and the work of \cite{tennie2023quantum, leyton2008quantum} use the forward Euler method to discretize the Lorenz system, there are two main differences. First, in this work, we simulate only a single time step, whereas \cite{tennie2023quantum} simulates several time steps (T) in a single iteration. The second big difference is the use of a collection of qubits in \cite{tennie2023quantum, leyton2008quantum} to give a mean-field approximation. Since we simulate only a single time step, we do not need a collection of qubits to avoid measurement errors that compound with time steps. 

On the other hand, \citet{berry2014high} has an efficient quantum procedure for linear differential equation, while in the present work, we consider a nonlinear system, so the approach here is not directly comparable with the results there.

The system \eqref{system3x}-\eqref{system3z} has nonlinear terms, specifically $xz$ and $xy$, that distinguish it significantly from the linear system presented in \cite{armaos2024quantum}. However, the present system is not as complex as the non-linear system presented in \cite{lloyd2020quantum}.\\
{To show the efficiency of the proposed method,  we compare the results of our proposed method and the Euler method, which is a non-linear approach,  to solve the Lorenz system. We use the same number of iterations, step size, and initial point.  We plotted this compression in Figure \ref{new-euler}.
\begin{figure*}[t!]
    \centering
        \includegraphics[height=2in]{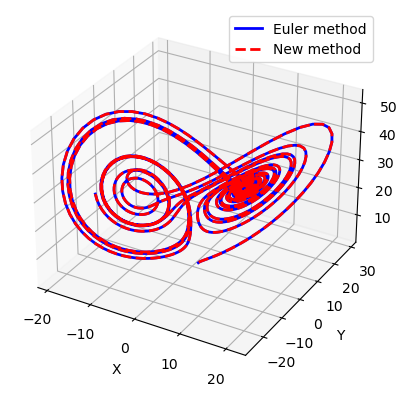}
        \caption{The starting point is $(1,1,1)$.}
        \label{new-euler}
\end{figure*}
Encouragingly, the solutions obtained from our proposed linear system approach are similar to the those from nonlinear Euler method. This result demonstrates that although our method is based on solving a linear system, it effectively captures the dynamics of the nonlinear Lorenz system. }
To solve (\ref{final_system}), we begin from an initial point and use iterative methods to find the optimal solution. While numerous classical methods are available \cite{butcher1996history}, 
 we will use a variational quantum algorithm called VQLS.

\section{Variational Quantum Linear Solver}
\label{sec:vqls}
\subsection{Overview}

The VQE is a sophisticated hybrid quantum-classical algorithm designed to determine the ground state energy of quantum systems, such as molecular structures. VQE operates by utilizing a parameterized quantum circuit to prepare an approximate quantum state, which is then measured to estimate the expectation value of the Hamiltonian, representing the system's energy. This energy estimation is subsequently fed into a classical optimization process that iteratively adjusts the circuit's parameters to minimize the energy, ultimately converging on the ground state \cite{peruzzo2014variational}.
The VQLS \cite{bravo2023variational} is a hybrid quantum algorithm for solving linear systems of equations using the variational principle; which minimizes the expectation value of the Hamiltonian of the system, with respect to a parameterized quantum circuit. Given $A$ and $b$, VQLS aims to prepare variationally a state $\ket{w}$ such that $A \ket{w} \sim \ket{b}.$ One of the main appeals of VQLS is that it can be implemented on near-term NISQ computers. Experimental studies on a limited set of test instances on Rigetti machines seem to offer evidence that VQLS scales linearly in the condition number $\kappa$ and $\log\frac{1}{\epsilon}$ where $\epsilon$ is the desired precision \cite{bravo2023variational}.

VQLS is a parameterized circuit. To find a good set of values for the parameters, classical optimization is used, which is computationally challenging. Estimating the values with a precision of $\pm 1/poly(n)$ is DQC1-hard, where $n$ represents the number of qubits. It is believed that classical algorithms cannot efficiently find precise values of the parameters because efficient simulation of DQC1 would lead to the collapse of the polynomial hierarchy to the second level. Consequently, there are doubts about the efficient classical simulation of VQLS.

In order to effectively use the VQLS algorithm, the input matrix $A$ must satisfy certain requirements. $A$ should be representable as a linear combination of unitaries, similar to how the Hamiltonian is represented in the variational quantum eigensolver as a linear combination of Pauli operators. The method provided by \cite{bravo2023variational} is based on Szegedy walks and efficiently decomposes a sparse matrix into a linear combination of unitaries. $A$ should also be sparse and well-conditioned, a finite $\kappa$. The norm of A should be bounded, $||A|| \le 1$. Lastly, the unitaries in the decomposition must be efficiently implementable. All these assumptions are satisfied by the matrix in \eqref{final_system}.

We assume that \( A \) can be expressed as a linear combination of unitary operators, such that 
\begin{equation}
    A = \sum_i c_i A_i 
\end{equation}
where \(A_i \) are the unitaries and \( c_i \) are complex coefficients. This representation effectively models a system Hamiltonian. 
Typically, the decomposition involves a linear combination of tensor products of the Identity and Pauli matrices. These gates are widely used due to their well-known properties and ease of implementation. The matrices representation of the gates are below.
\begin{eqnarray}
I = \begin{pmatrix} 1 & 0 \\ 0 & 1 \end{pmatrix}, \quad 
X = \begin{pmatrix} 0 & 1 \\ 1 & 0 \end{pmatrix}, \quad 
Y = \begin{pmatrix} 0 & -i \\ i & 0 \end{pmatrix}, \quad 
Z = \begin{pmatrix} 1 & 0 \\ 0 & -1 \end{pmatrix}.
\end{eqnarray}

We use a recently proposed algorithm to decompose a square real symmetric matrix of any size into a tensor product of Pauli spin matrices for all application matrices discussed. This algorithm, which is detailed in  \cite{pesce2021h2zixy}, is available in Pennylane and has been utilized to generate decompositions for stiffness matrices of general sizes commonly encountered in discrete finite-element methods. We do not use the decomposition algorithm from \cite{bravo2023variational}.

\subsection{Cost function}
Two types of cost functions have been introduced for the VQLS method: local cost functions and global cost functions. We describe the cost functions and highlight their features. 

The residual-based cost function is given by
\begin{equation}
 C({\theta}) = \min_{\theta}\|A |\psi({\theta})\rangle -|{b}\rangle\|^2 = \min_{\theta}\langle (A |\psi({\theta})\rangle - |{b}\rangle) ^{\dagger} (A |\psi({\theta})\rangle - |{b}\rangle) \rangle.
\end{equation}

This cost function can be viewed as 
the expectation value of an effective Hamiltonian which  is defined as follows \cite{bravo2023variational}:
\begin{equation}\label{Hamil}
    H_G=A^{\dagger}\left(I-|b\rangle\langle b|\right)A.
\end{equation}
Therefore, we can write the cost function 
associated with Hamiltonian $H_G$ as:
\begin{equation}\label{global-cost}
    C_G(\theta)=\langle \psi(\theta)|H_G|\psi(\theta)\rangle \,.
\end{equation}


\subsection{Ansatz}
{In variational algorithms, an Ansatz refers to an assumed initial form for the quantum state. This Ansatz is typically represented by a parameterized quantum circuit, which is used to prepare a trial state that can be optimized.}
The Strongly Entangling Layers (SEL) Ansatz is a type of parameterized quantum circuit used in variational quantum algorithms, such as the VQE and Quantum Approximate Optimization Algorithm (QAOA). This Ansatz is designed to introduce a high degree of entanglement between qubits while maintaining a relatively simple and regular structure, making it a popular choice for many quantum applications. Let's describe the structure of the SEL Ansatz. 

Each SEL Ansatz consists of layered circuits composed of rotation gates and entanglement operations:
\begin{itemize}
    \item Parameterized Rotation Gate:
   Each qubit undergoes a parameterized rotation, typically represented as $R(\alpha,\beta,\gamma)$. This gate can be a combination of rotations around different axes, such as $R_X$, $R_Y$, and $R_Z$.
\item Entangling Operations:
   Following the rotation gates, to entangle the qubits, a series of controlled-NOT (CNOT) gates are applied. The pattern of these CNOT gates may vary, but they generally ensure that every qubit is entangled with at least one other qubit in the layer.
\item  Layered Structure:
   To enhance the expressiveness of the Ansatz, multiple layers of the above combination are stacked. Each layer applies a new set of parameterized rotation gates $R(\alpha,\beta,\gamma)$ 
\begin{equation}
\begin{split}R(\alpha,\beta,\gamma) = R_Z(\gamma)R_Y(\beta)R_Z(\alpha)= \begin{bmatrix}
e^{-i(\alpha+\gamma)/2}\cos(\beta/2) & -e^{i(\alpha-\gamma)/2}\sin(\beta/2) \\
e^{-i(\alpha-\gamma)/2}\sin(\beta/2) & e^{i(\alpha+\gamma)/2}\cos(\beta/2)
\end{bmatrix}.\end{split}    
\end{equation}
followed by entangling CNOT gates.
\end{itemize}

\begin{figure*}[!ht]
\centering
  \includegraphics[width=1\textwidth]{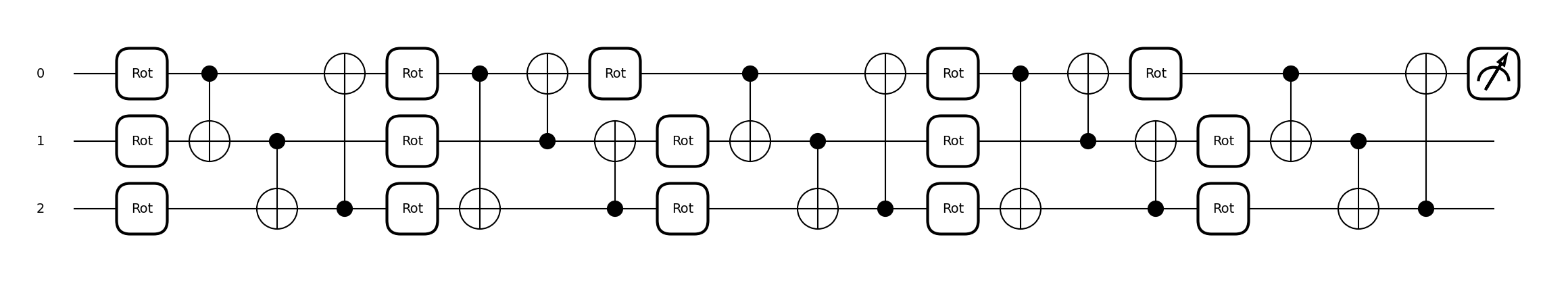}
  \caption{Five layer Ansatz used in the VQLS algorithm. 
  }
  \label{fig:lor}
\end{figure*}



\section{Algorithm} \label{sec:algorithm}

To ease the notation, since the focus will be fully on the nonlinear system (\ref{final_system}), from here on we indicate with $A$ the matrix $A_{NL}$, with $w$ the vector $W$ and with $b$ the vector $b_{NL}$, see Eqs. \eqref{matrixnonlineq}-\eqref{final_system}.

\begin{algorithm}
\caption{Variational Quantum Linear Solver \cite{bravo2023variational}}
\begin{algorithmic}[1]
\Require Matrix $A$, vector $b$, number of layers  $num\_layers$,  maximum iterations $max\_iterations$, convergence tolerance $conv\_tol$, step size $stepsize$
\Ensure Optimized parameters
\State $A^\dagger \gets$ Hermitian conjugate of $A$
\State $b\_norm \gets b / \|b\|$
\State $P_b \gets b\_norm \otimes b\_norm^T$
\State $I \gets$ Identity matrix of size $A$
\State $H_G \gets A^\dagger \cdot (I - P_b) \cdot A$
\State Define cost function $C_G$
\State $num\_qubits \gets \lceil \log_2(\text{size of } A) \rceil$
\State Define quantum device with $num\_qubits$
\State Define Ansatz with $num\_layers$
\State Initialize optimizer with $stepsize$
\State Initialize random parameters for Ansatz
\For{$it \gets 1$ to $max\_iterations$}
    \State Compute gradient 
    \State Update parameters and compute the cost function
    \State Check stop condition
\EndFor
\State \Return optimized parameters 
\State  Extract the solution
\end{algorithmic}
\label{Algo1}
\end{algorithm}

Algorithm \ref{Algo1} outlines the VQLS method for solving a system of linear equations. The algorithm starts with the following inputs: a specified matrix \( A \), vector \( b \), the number of layers, the maximum number of iterations, the convergence tolerance, and a fixed step size for the optimization process.

In lines 1-6 of the algorithm, the cost function is defined based on the matrix \( A \) and the vector \( b \). After defining \( H_G \), the cost function is evaluated using the expectation value of $H_G$. The number of qubits and the Ansatz are specified. Initially, the parameter \( \theta \) is assigned a random value. The rest of the algorithm is the classic gradient descent loop. The algorithm then computes the cost function and gradient within the for loop and updates the parameters \( \theta \). This process repeats until the stopping condition is met. Once the optimal value for the parameter \( \theta \) is found, the solution for the system \( A w = b \) is determined. It is important to note that the solution must satisfy \( A w - b = 0 \) to be considered valid.
\section{Numerical Results}
\label{sec:simul}

We present the results of implementing the quantum Algorithm 1 and classical approaches to solve the nonlinear Lorenz system  \eqref{final_system}. We use the following values of the initial parameters:
\begin{itemize}
    \item Number of layers: The number of layers is a crucial parameter for achieving optimal parameter values. We ran the algorithm with the same initialization and plotted these behaviors of the cost functions in Figure \ref{fig:lor} as a function of the number of layers. Three or more layers appear to be {\it necessary} to compute a good set of values for the parameters.
\begin{figure*}[!ht]
\centering
  \includegraphics[width=0.45\textwidth]{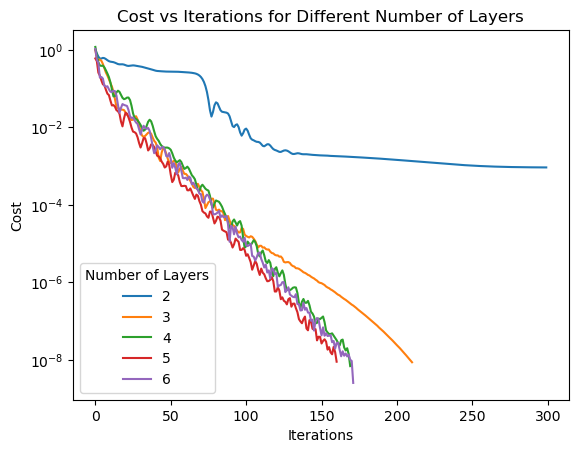}
  \caption{Expectation value as a function of the number of layers.}
  \label{fig:lor}
\end{figure*}
    
    \item Stop condition: We used two different criteria for the stop condition: the maximum number of iterations and convergence tolerance are given by:
\begin{equation*}
\text{Max-iter}=200,\quad \epsilon=1e-08 \,,
\end{equation*}
respectively.

    \item Initialization:
    The initial value of the parameter $\theta$ is chosen randomly. 
    \item Cost function:
    We consider the following cost function
    \begin{equation*}
        C_G(\theta)=\langle \psi(\theta)|H_G|\psi(\theta)\rangle
    \end{equation*}    
    
    \item Number of qubits:
Error-free qubits are an important quantum resource. In our implementation, three qubits are needed, which is a very low resource requirement compared, to algorithms based on HHL in which an extra register is needed to store the eigenvalue computed during the phase inversion process. For greater precision, a larger sized quantum register is needed to store the eigenvalues. In general, the VQLS algorithm needs \(\log n\) qubits, where \(n\) is the size of the matrix \(A\).

\end{itemize}
\subsection{Solution}
Quantum states have unit norms.  Thus, to find the exact solution that satisfies the system of equations \(A w = b\) where $b$ is not of unit length, we need to scale the state vector. Once the optimal parameters are obtained using the classical optimization, we execute the Ansatz to obtain the final state. Now, this state can only be measured, and we make an assumption that we can reconstruct $w$ from $\ket{w}$ efficiently. Next, we seek a coefficient \(C\) such that the solution of the VQLS can be scaled by $C$ to match the exact solution $w = A^{-1}b$.
To determine the value of \(C\), we divide \(\|b\|\) by \(\|A\Tilde{w}\|\). By multiplying this factor by the solution, we obtain the exact solution.

\subsection{Handling the Sign of the Solution}
We know that if \(w\) is an eigenvector of the matrix \(A\), then \(-w\) is also an eigenvector of \(A\). To obtain the correct solution, we check the distance between \(A w\) and \(b\) for the \(i\)-th element in the solution. If this distance is less than the absolute value of \(A w(i)\), we do not need to multiply by $-1$. Otherwise, we must multiply by $-1$ to ensure the correct solution.

\subsection{Condition number $\kappa$}

A well-known measure of how ill-conditioned is a non-singular matrix $A$ is the condition number $\kappa(A) = ||A^{-1}|| * ||A||$ where $|| \circ ||$ is any norm. When $2$-norm is used the condition number is the ratio of largest to smallest eigenvalue. The condition number of the input matrix \( A \) significantly affects the performance of the VQLS algorithm; the running time is linear in $\kappa$. Specifically, for the matrix \( A \) defined by (\ref{final_system}), the condition number depends largely on the step size parameter \( h \). As \( h \) gets closer to 1, the condition number of \( A \) increases exponentially in general. On the other hand, the lowest condition number is achieved as \( h \) tends towards zero. As mentioned in \cite{bravo2023variational}, the VQLS algorithm performs optimally when the condition number is small. In our study, we use \( h = 0.01 \) to achieve better performance. Figure \ref{cond} shows the relationship between the condition number of the matrix \( A \) and the value of \( h \). Even for large values of step size ($\simeq 0.1$), the condition number is bounded by 70.

\begin{figure*}[!ht]
\centering
  \includegraphics[width=0.45\textwidth]{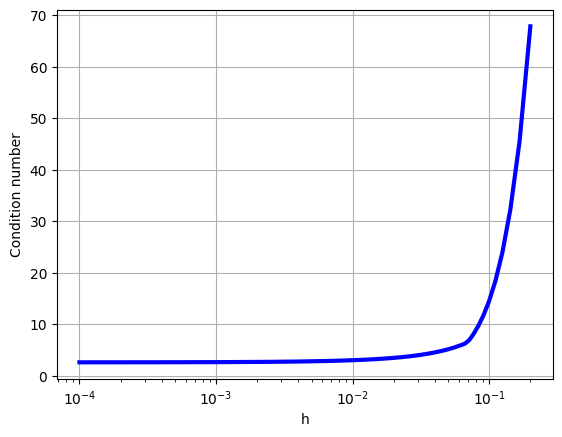}
  \caption{The relationship between the condition number of matrix \( A \) and the value of \( h \).}
  \label{cond}
\end{figure*}

\subsection{Starting point}

To solve system (\ref{final_system}) using the VQLS, we begin by initializing the algorithm with the following values:
\begin{itemize}
    \item  $b=(1,-2,4,0,0,0,4,-2)$, $h=5*10^{-3}$, $ \sigma =10$,  $\rho = 28$,  $\beta = \frac{8}{3}$, and $T=2000$.
    
    \item $b=(10^{(-16)},-10^{(-16)},10^{(-16)},0,0,0,0,0)$, $h=10^{-3}$, $ \sigma =10$,  $\rho = 13.92655742$, and $\beta = \frac{8}{3}$, and $T=10000$.
\end{itemize}
%

\subsection{Error Analysis}
\label{sec:error}

To compare the trajectories calculated using the classical and quantum algorithms, as outlined in the methods section, we can analyze them point by point. For each time step, we calculate the difference in the coordinate positions obtained from the two methods and normalize them using the formula provided next. Let $w^c_n = (x_n^c,y_n^c,z_n^c,)$ and $w^q_n = (x_n^q,y_n^q,z_n^q,)$ represent the points computed in the classical and quantum algorithms, respectively, at iteration $I = n$. The relative error is defined as:
\begin{equation}
    \frac{|x^c_n-x^q_n|+|y^c_n-y^q_n|+|z^c_n-z^q_n|}{1+|x^c_n|+|y^c_n|+|z^c_n|} \,.
\end{equation}
The relative error for the two algorithms after 500 iterations is illustrated in Figure \ref{error-quantum}. This shows that the trajectories calculated by the quantum method proposed in this paper closely match the trajectory computed using a classical computer. 
\begin{figure*}[!ht]
\centering
\includegraphics[width=0.4\textwidth]{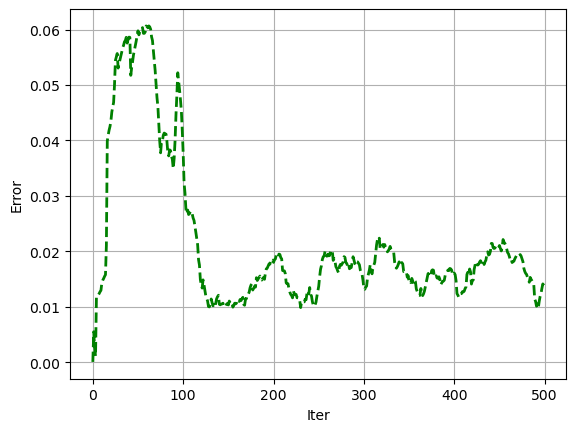}
  \caption{The relative-error of the quantum and classical methods for 500 iterations with a step size of 0.001. The average error is order of magnitude larger than the step size.}
  \label{error-quantum}
\end{figure*}

However, analyzing relative error alone does not provide insights into the impact of step size on error. Therefore, we conducted a more comprehensive analysis using Richardson's extrapolation method.

For a deeper analysis of errors, we consider the error in computing the gradients $\dot{x}, \dot{y}, \dot{z}$. This error serves as a good proxy for the error in computing the trajectories, as the only error that occurs in each iterative step is an error in the gradient computation. We make the strong assumption that errors arising from quantum uncertainty are negligible.

Let $\dot{x_h}$ be the numerically computed gradient as a function of the step size $h$, where $h > 0$ represents a positive number close to zero. The numerical gradient $\dot{x_h}$ can be expressed as the actual gradient $\dot{x}$ plus some error $E(h)$, which is a function of $h$:
\begin{equation}
    \dot{x}_h=\dot{x} +E(h) \,,
\end{equation}
where \( E(h) \) is the error associated with the forward difference method. The error term \( E(h) \) for the forward difference method is
\begin{eqnarray}
   E(h) = c \cdot h + O(h^2) \,,
\end{eqnarray}
where $c$ is a constant. This relationship is true for all step sizes $h$. Therefore, the numerically computed derivatives for steps size $h, 2h$ can be written as 
\begin{eqnarray}
     \dot{x}_h=\dot{x} +ch+O(h^2)\\
    \dot{x}_{2h}=\dot{x} +c(2h)+O(h^2) \,.
\end{eqnarray}
We obtain the error for the step size \(h\) by subtracting the second equation from the first:
\begin{equation}
\label{eq:err}
  \dot{x}_{2h} -  \dot{x}_h= ch + O(h^2) = E(h) \,.
\end{equation}

This intuitive calculation can be formalized, as shown in \cite{richardson1911ix}. At each time step $n$, we use the numerical gradients computed at step sizes $h,2h$ to bound the error in the computation. The initial point for both the step sizes is point $(x_{n-1}, y_{n-1}, z_{n-1})$. Similarly, we can obtain the error in computation of $\dot{y}, \dot{z}$. The numerical derivatives are computed in two ways: classical and using a quantum algorithm. We compute the error for both methods. 

We use \eqref{eq:err} to estimate the error for algorithms (both classical and quantum) as a function of the step size. We use the error measure to quantify the error of the two approaches individually. We also use the error to do a comparative analysis. VQLS algorithm takes
{
\subsection{Error for VQLS Method}
Here, we briefly review some key concepts related to errors in the VQLS method \cite{bravo2023variational}.
For the VQLS algorithm, the deviation between observable expectation values for the approximate solution $|x(\alpha_{\text{opt}})\rangle$ and the true solution $|x_0\rangle$ can be upper bounded based on the value of the cost function. Specifically, the error tolerance $\epsilon$ can be set before running the algorithm, where $\epsilon$ is defined as the trace distance between the exact and approximate solutions:
\begin{equation}\label{2}
    \epsilon = \frac{1}{2} \text{Tr} \left( \left\| |x_0\rangle \langle x_0| - |x(\alpha_{\text{opt}})\rangle \langle x(\alpha_{\text{opt}})| \right\| \right) 
\end{equation}
Moreover, it has been proven that the cost function  satisfies the relation \cite{bravo2023variational}:
\begin{equation}
    C_G \geq \frac{\epsilon^2}{\kappa^2}
\end{equation}
where $\kappa$ is the condition number of the input matrix, and $C_G$ is the cost function defined by (\ref{global-cost}). By using the above formulas, we can observe that the difference between the exact solution and the quantum solution is related to $\kappa^2$ multiplied by the value of the objective function. This shows that the error between the quantum solution and the exact solution is inherently connected to the condition number of the matrix and the chosen cost function, allowing us to control and quantify the error before running the algorithm.}

{
        Several noise mitigation strategies are particularly well-suited to enhance the robustness of VQLS on NISQ devices due to the method's variational structure and we can apply these methods. Techniques such as Zero-Noise Extrapolation (ZNE) and Measurement Error Mitigation (MEM) allow the VQLS to reduce the impact of noise without increasing the quantum circuit depth, making them practical for the shallow circuits typically used in variational algorithms \cite{temme2017error, kandala2019error}. Additionally, the Variational Error Suppression (VES) technique can adaptively adjust the parameters within VQLS to minimize noise effects, enhancing solution accuracy even on noisy devices \cite{McClean2017, Cerezo2021}. Together, these approaches demonstrate VQLS's practical suitability for today’s quantum hardware while mitigating the effects of common noise sources.
}

\subsection{Results and Discussion}
\label{sec:results}

The VQLS circuit has a short depth (20). Short depth circuits can be implemented in NISQ computers because these circuits require fewer quantum gates and operations to execute. NISQ computers are currently limited by noise and errors, which makes long computations and complex circuits prone to errors. Short depth circuits require fewer operations, reducing the likelihood of errors and increasing the chances of successfully executing the computation on current NISQ hardware. Short depth circuits also offer better error mitigation \cite{temme2017error}. 

One of the critical resources used by any quantum algorithm is the number of qubits, which significantly impacts the simulation time. The algorithm proposed here to solve the Lorenz system using the VQLS method requires only three fault-tolerant logical qubits. 
However, since the simulations were performed using a classical computer, the execution time is slightly longer than that of classical methods. However, it should be noted that this simulation time is insignificant compared to the simulation of the HHL method.
We coded the VQLS and a classical algorithm.

\begin{figure*}[t!]
    \centering
    \begin{subfigure}[t]{0.45\textwidth}
        \centering
        \includegraphics[height=2in]{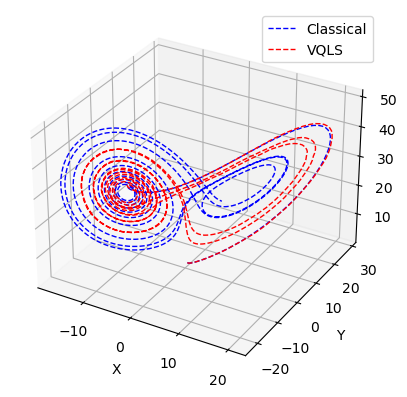}
        \caption{Comparison of classical and quantum results for the first 2000 iterations.}
    \end{subfigure}%
    ~ 
    \begin{subfigure}[t]{0.45\textwidth}
        \centering
        \includegraphics[height=2in]{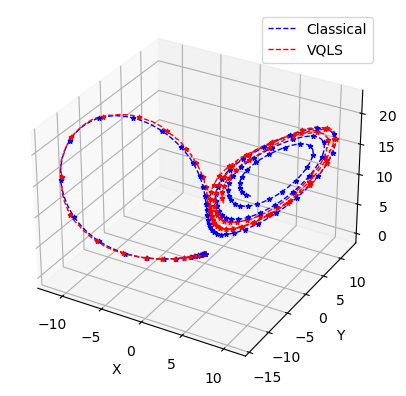}
        
        \caption{Trajectories computed by classical and quantum simulations. The initial point is $(1e^{-16},-1e^{-16},1.e^{-16})$ and 10000 timesteps are shown. The same attractor is discovered by both methods.}
    \end{subfigure}
    
    \caption{Trajectories computed using classical and Quantum methods}
    \label{fig:clas-vqls}
\end{figure*}

An important observation is that, although we solved a linear system of equations, the results resemble those of a nonlinear system. The new system has nonlinear product terms, so it simulates the nonlinear system of differential equations well in low dimensions.

For two different start points, the trajectories computed using the classical and the VQLS methods are shown in Fig. \ref{fig:clas-vqls}. Both approaches compute similar trajectories. The chaotic system appears to have more error compared to the attractor.

We use equation \eqref{eq:err} to estimate the error for algorithms (both classical and quantum) as a function of the step size. We use the error measure to quantify the error of the two approaches individually. We also use the error to do a comparative analysis. VQLS algorithm takes considerable time to simulate; therefore, only a limited comparative study is conducted.  

First, we examine the computation shown on the left side of Figure \ref{fig:chaotic}. The start point is $(1,-2,4)$. Figure \ref{fig:classerr} (a) shows the error in the computation of $\dot{x}, \dot{y}, \dot{z}$ in each iteration and the total error (defined as the sum of individual errors). It is evident that the equation system \eqref{final_system} solved using the classical approach has good accuracy. The average total error is the same for a step size of $10^{-3}$. The error decreases as the number of iterations increases.

Next, we study the classical error in the computation of the blue curve shown in Figure \ref{fig:start}. Figure \ref{fig:classerr} (b) plots the individual and the total error as a function of the iteration for a step size of $10^{-3}$. For the first 200 iterations, the total error is less than $10^{-16}$. Initially, the error is insignificant, but for the last half of the computation, the error is significant, close to 100 times more than the step size. The average error is still comparable to the step size. 

\begin{figure*}[t!]
    \centering
    \begin{subfigure}[t]{0.45\textwidth}
        \centering
        \includegraphics[height=2in]{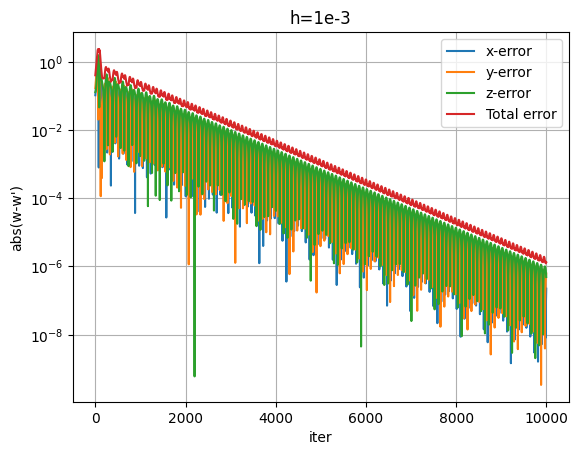}
        \caption{The error for the trajectory computed in Figure \ref{fig:chaotic}.}
    \end{subfigure}%
    ~ 
    \begin{subfigure}[t]{0.45\textwidth}
        \centering
        \includegraphics[height=2in]{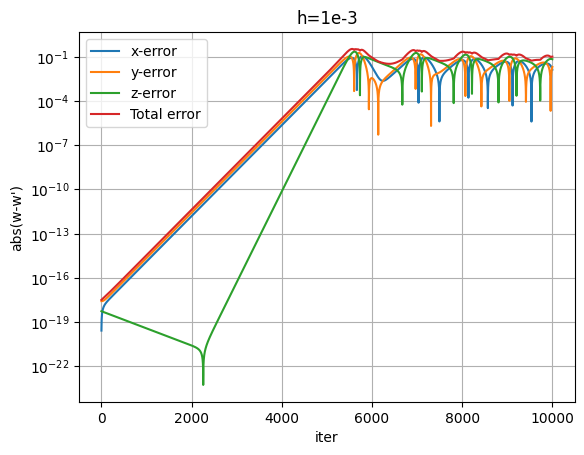}
        
        \caption{The error for the trajectory computed in Figure \ref{fig:start}.}
    \end{subfigure}
    
    \caption{The error for classical computation as given by equation \eqref{eq:err}.}
    \label{fig:classerr}
\end{figure*}

The total error in the two figures is of the order of the step size, so we infer that the system given by \eqref{final_system} is a good model. 

Next, we compare the error of the quantum method with the classical method. Since VQLS take a long time to simulate, we limit the experiments to two start points, the number of iterations to 200 in the preliminary results reported here. The individual errors and the total error in the computation of $\dot{x}, \dot{y}, \dot{z}$ for the trajectories shown in Figure \ref{fig:clas-vqls} (a) and Figure \ref{fig:clas-vqls} (b) are shown in Figure \ref{fig:vqls_classical_error} (a), (b) respectively. It appears the error of the hybrid quantum method is comparable. 

\begin{figure*}[t!]
    \centering
    \begin{subfigure}{0.45\textwidth}
        \centering
        \includegraphics[height=2in]{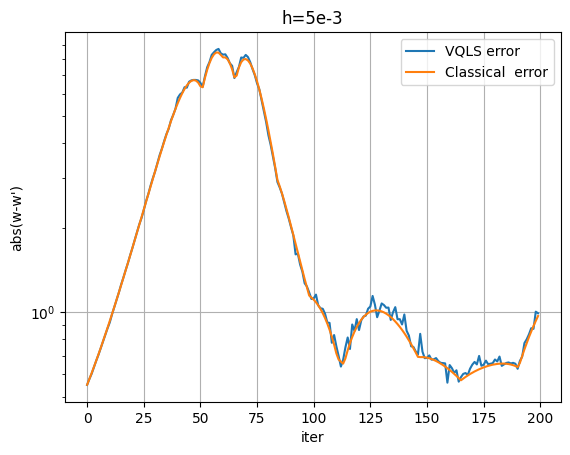}
        \caption{Errors for the trajectories computed in Figure \ref{fig:clas-vqls} (a).}
    \end{subfigure}%
    ~ 
    \begin{subfigure}{0.45\textwidth}
        \centering
        \includegraphics[height=2in]{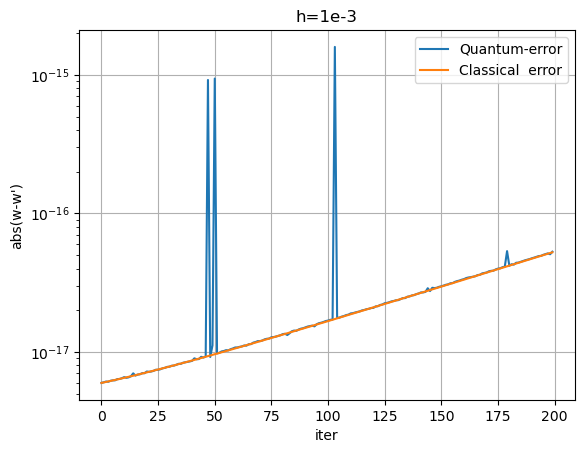}
        
        \caption{The error for the trajectory computed in Figure \ref{fig:clas-vqls} (b).}
    \end{subfigure}
    
    \caption{Classical Error v/s Quantum Error.}
    \label{fig:vqls_classical_error}
\end{figure*}



\subsection{Limitations}

The VQLS-based method is capable of handling nonlinearities. However, it does not offer the same exponential advantage as HHL when dealing with a "giant" linear transformation. The effectiveness of the VQLS method depends upon parameter values computed using a classical optimization approach, which are also sensitive to the initial points. The method relies on strong assumptions such as the requirement for exact preparation of initial states, noise-free evolution, and the ability to recover the final answer with the desired precision. The impact of quantum uncertainty on trajectories is not well understood; while we have shown that trajectories computed by classical and quantum algorithms are "close", it is conceivable that quantum uncertainty can lead to completely different trajectories from the classical ones, which requires further examination. VQLS is a hybrid algorithm, and simulating it on a classical computer is more time-consuming than computing the inverse of a small matrix. At present, there is no computational advantage for nonlinear difference equations involving few variables. 

{It is worth noting that the VQLS algorithm proposed for solving the Lorenz system aligns with existing implementations of quantum linear solvers for near-term quantum devices \cite{bravo2023variational}. In that study, a 10-qubit implementation of VQLS was performed on Rigetti’s Aspen-4 quantum computer to solve a Quantum Linear System Problem (QLSP) with a 1024 × 1024 matrix. This experiment utilized a specifically tailored ansatz incorporating $R_y(\alpha_i)$ gates and computed the cost function by expanding the Hamiltonian in terms of Pauli operators to meet hardware constraints. The findings indicated that the cost function was minimized effectively on quantum hardware, closely matching simulated results and verifying that the solution to the linear system was obtained. This research,  supports the feasibility of implementing our proposed method on current quantum devices.}
{\subsection{Matrix Formulation for High-Dimensional Systems}
To extend the proposed method to solve larger nonlinear systems of differential equations, we first need to ensure that the matrix $A_{NL}$  defined by (\ref{matrixnonlineq}) is compatible with quantum algorithms, which often require matrix dimensions to be powers of 2. If the original matrix $A_{NL}$ does not meet this requirement, we can pad it by adding an identity matrix or zero elements to reach the nearest  $2^n \times 2^n$  dimension. This allows the system to be compatible with VQLS and enables efficient processing on quantum hardware. 
\\
Consider a general nonlinear system of ordinary differential equations  given by:
\begin{eqnarray}
    \frac{d\mathbf{u}}{dt} = \mathbf{f}(\mathbf{u}(t)),
\end{eqnarray}
where $\mathbf{u}(t) = [u_1(t), u_2(t), \ldots, u_N(t)]^\top$  is an  $N$-dimensional vector of state variables, and  $\mathbf{f}(\mathbf{u}) = [f_1(\mathbf{u}), f_2(\mathbf{u}), \ldots, f_N(\mathbf{u})]^\top$ represents the nonlinear terms.
\\
Using a timestep  $h$, the discretized form at each time step  $n$ can be written as:
\begin{eqnarray}
  \mathbf{u}_{n+1} = \mathbf{u}_n + h \mathbf{f}(\mathbf{u}_n),  
\end{eqnarray}
which we rearrange as:
\begin{eqnarray}\label{non-format}
    \mathbf{u}_{n+1} - \mathbf{u}_n - h \mathbf{f}(\mathbf{u}_n) = 0.
\end{eqnarray}
Now, we can rewrite the (\ref{non-format}) into matrix form as:
\begin{equation}\label{gen}
    A_{NL} W = b_{NL},
\end{equation}
where:
$A_{NL}$  is an $(N + M) \times (N + M)$ matrix (or nearest  $2^n \times 2^n$  size if padding is applied), $ W$  is an  $(N + M)$ -dimensional vector containing state variables at the current and next time steps, and
$ b_{NL}$ denotes an  $(N + M)$ -dimensional vector with known values.
Now, we can apply the VQLS to solve the linear system given by (\ref{gen}).}

\section{Conclusion}

This is an exploratory study on the use of quantum algorithms for studying complex systems. We showed how a variational quantum algorithm can be used to solve the Lorenz system. We perform a comparative error analysis of the quantum method with the classical method. The quantum method is found to be reliable in the simulations. The method has limitations, such as dependence on classical optimization, read-out, read-in capabilities of quantum systems, and noise in quantum computation.

\section*{Abbreviations}{
The following abbreviations are used in this manuscript:\\

\noindent 
\begin{tabular}{@{}ll}
QRAM &  Quantum random-access memory\\
ODEs & Ordinary differential equations\\
HHL & Harrow-Hassidim-Lloyd\\
QPCA & Quantum principal component analysis\\
VQE &  Variational quantum eigensolver \\
VQLS & Variational quantum linear solver\\
NISQ & Noisy intermediate-scale quantum\\
SEL & Strongly entangling layers\\
QAOA & Quantum approximate optimization algorithm\\
CNOT & Controlled-NOT\\
VES &Variational Error Suppression\\
ZNE&Zero-Noise Extrapolation\\
MEM&  Measurement Error Mitigation\\
QLSP&Quantum Linear System Problem\\
\end{tabular}
}

\section*{Acknowledgments}{We would like to thank Sarah Alibhai, Maxim Ciobanu, Brianna Huff, Jaanatul Moawa, David Peters, Sreyas Saminathan, Md Mohsin Uddin,  Fabliha Zahin for discussions in the initial stages of this work.

A.I. gladly acknowledges support from Charles University Research Center (UNCE 24/SCI/016).}

\bibliographystyle{unsrt}  

\end{document}